\documentclass{jfm}
\usepackage{graphicx}
\usepackage{epstopdf, epsfig}
\usepackage{color}

\shorttitle{Investigation of nonlocal data-driven methods for SGS modelling in LES}
\shortauthor{B. Liu, H. Yu, H. Huang and X.-Y. Lu}

\title{Investigation of nonlocal data-driven methods for subgrid-scale stress modelling in large eddy simulation}

\author{Bo Liu\aff{1},
  Huiyang Yu\aff{1},
  Haibo Huang\aff{1},
 \and Xi-Yun Lu\aff{1}
 \corresp{\email{xlu@ustc.edu.cn}}}

\affiliation{\aff{1}Department of Modern Mechanics, University of
Science and Technology of China, Hefei, Anhui 230026, China}

\begin{document}

\maketitle

\begin{abstract}
A nonlocal subgrid-scale stress (SGS) model is developed based on the convolution neural network (CNN), a powerful supervised data-driven approach. The CNN is an ideal approach to naturally consider nonlocal
spatial information in prediction due to its wide receptive field.
The CNN-based models used here only take primitive flow variables
as input, then the flow features are automatically extracted without any $priori$
guidance. The nonlocal models trained by direct numerical
simulation (DNS) data of a turbulent channel flow at $Re_{\tau}=178$ are accessed in both the $priori$ and $posteriori$ test, providing physically reasonable flow statistics (like mean velocity and velocity fluctuations) closing to the DNS results even when extrapolating to a higher Reynolds number $Re_{\tau}=600$. In our model, the backscatter is also predicted well and the numerical simulation is stable. The nonlocal models outperform
local data-driven models like artificial neural network and some SGS
models, e.g. the Smagorinsky model in actual large eddy simulation
(LES). The model is also robust since stable solutions can be
obtained when examining the grid resolution from one-half to double
of the spatial resolution used in training. We also investigate the
influence of receptive fields and suggest using the two-point correlation
analysis as a quantitative method to guide the design of
nonlocal physical models. To facilitate the combination of machine
learning (ML) algorithms to computational fluid dynamics (CFD),
a novel heterogeneous ML-CFD framework is proposed.
The present study provides the effective data-driven nonlocal
methods for SGS modelling in the LES of complex anisotropic turbulent flows.

\end{abstract}


\section{Introduction}\label{sec:intro}

Large eddy simulation (LES) is a powerful tool for turbulence simulation.
The flow fields in LES are decomposed into
the resolved and unresolved parts. The resolved large-scale
turbulent motions are directly solved from the Navier-Stokes equations,
while the unresolved part is modelled by subgrid-scale (SGS) stress
models. In the recent decades, a variety of SGS models are developed to
establish mappings from the resolved flow variables to the SGS
stress tensor. Based on the Boussinesq hypothesis, the first
SGS model proposed by \cite{smagorinsky1963}
represents that the SGS stress is linearly related to the resolved strain rate tensor.
This model is simple and effective but purely
dissipative since no backward energy transfer from the subgrid-scale to
the resolved scale is allowed. The dynamic models
are motivated by the defect of the Smagorinsky model \citep{germano1991dynamic,lilly1992proposed,meneveau1996lagrangian}.
The local value of Smagorinsky coefficient is dynamically determined
in different flow regimes and the backscatter can be
predicted. In reality, the backward energy transfer from residual
motions to the resolved flow field is clipped sometimes by setting
the negative eddy viscosity $\nu_t$ to zero considering the
numerical stability \citep{zang1993dynamic,vreman1997large}. The
dynamic models are generally time-consuming because of the
additional filtering operation. On the other hand, the gradient
model \citep{clark1979evaluation,liu1994properties} and the
similarity model \citep{bardina1983improved,domaradzki1997subgrid}
are not sufficiently dissipative and prone to be numerically
unstable in actual LES.

Recently, data-driven approaches are introduced to LES
modelling \citep{yang2019predictive,sirignano2020dpm,yuan2020deconvolutional,subel2021data}.
\cite{gamahara2017} attempted to construct SGS stress models from
filtered DNS (fDNS) data of turbulent channel flow using the
artificial neural network (ANN). In their work, the strain tensor,
rotation tensor, and the distance from the wall are selected as the
input features of ANN, while the performance of ANN shows no
advantage over the Smagorinsky model in the $posteriori$ test.
\cite{wang2018} compared the performance of two machine learning
(ML) algorithms, i.e. random forests and ANN, they found that ANN is
better in SGS modelling, while random forests are helpful for input
feature selection. Besides, \cite{maulik2018data} used the blind
deconvolution method to recover unfiltered variables from the
corresponding filtered ones. In compressible flow field,
\cite{xie2019artificial} applied an ANN mixed model to predict the
SGS stress and the SGS heat flux of compressible isotropic
turbulence.

\begin{figure}
  \centerline{\includegraphics[width=0.9\textwidth]{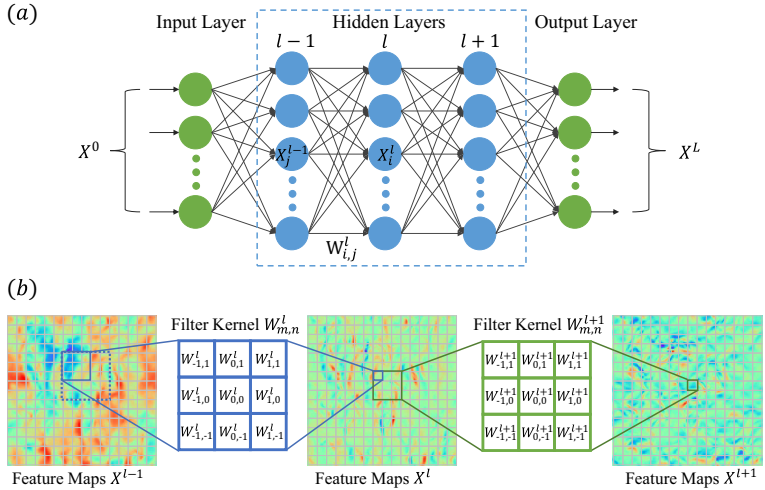}}
\caption{(a) A schematic of ANN and (b) CNN with
two-dimensional convolutional kernels whose kernel size is
$3\times3$. The rectangular boxes of 3-by-3 grid size in the feature
maps are the receptive fields where the filter kernels work on. The
1-by-1 grid size boxes are the corresponding output. The activation
functions and bias terms are omitted for clarity.}
\label{fig:ann_cnn}
\end{figure}

Most models mentioned above are working in a pointwise manner, i.e.
only local information is considered in prediction. Some researchers
\citep{maulik2019subgrid,xie2019modeling} choose a multiple points
stencil as input of ANN to include non-local spatial information.
The research of \cite{park2021} shows the model with multiple grid
points type input can obtain larger correlation coefficients in
contrast to the one with single grid point type input in the $priori$
test but suffers numerically unstable in $posteriori$ simulation. In
the framework of ANN as shown in figure \ref{fig:ann_cnn}(a), the number of
input features multiplied with the increase of the number of stretch
grid points, which leads to larger networks and lower efficiency.
Besides, feature engineering is a common issue for ANN-based models
no matter how many grid points are included.

Previous researches have shown that input feature selection is
crucial to the success of ANN-based models and extra spatial
information helps improve the model's performance. In the machine
learning field, the convolutional neural network (CNN) is another
powerful tool that is good at feature extraction and naturally
contains spatial topology information. Using CNN, we can simply use
primitive variables as input, and let the machine learn a model from
DNS data automatically without any assumption. Apart from that, CNN
has a much wider scope in contrast to ANN with stencil-type input.
As depicted in figure \ref{fig:ann_cnn}(b), the scope size
increases with the deepening of the network, any prediction made on
the last layer is based on the information from a block of the first
layer rather than a point.

Resulting from the above-mentioned advantages, CNN has started to be
used in the subgrid-scale modelling
\citep{bolton2019applications,zanna2020data}. \cite{beck2019deep}
constructed a mapping from coarse grid quantities to SGS force with
CNN in decaying homogeneous isotropic turbulence. Their $priori$ test
result indicates CNN-based models can get better results than
ANN-based ones. \cite{pawar2020priori} conducted a $priori$ analysis
on different data-driven parameterizations (i.e. ANN with
point-to-point mapping, ANN with neighbouring stencil data mapping
and CNN) for LES of two-dimensional Kraichnan turbulence. They found
that CNN can provide the most accurate predictions with less
computational overhead. Although these $priori$ tests have demonstrated
the potential of CNN-based models, further test in practical
simulations is absent. Some obstacles prevent CNN from being used in
actual LES (the $posteriori$ test), mainly about efficiency and
manoeuvrability. Firstly, CNN would be quite slow when serially
working on the CPU. Only when running in a highly parallelized way,
it shows a speed advantage. However, most CFD programs are written
for CPU which cannot meet the demand of CNN. Secondly, Python is
the most widely used computational language in constructing
machine-learning models like CNN, various Python open-source
libraries make programming convenient. Nevertheless, CFD codes are
mainly written in Fortran or C++, so mixed-language programming may
be an irritating issue when embedding CNN models to realistic LES.
One solution is writing both the ML and CFD codes in the same
programming language, but it may take a lot of time to rewrite the
source code and the efficiency problem still exists. Another
solution is using mix-language programming, while the CFD process
and ML process have to run on the same device or machine in this
way.

In this work, we aim to develop a concise, non-local, efficient SGS
stress model using CNN. To overcome the obstacles mentioned above,
we propose the heterogeneous Machine-learning CFD (HML-CFD)
framework based on inter-process communication technology which
breaks the limitation of programming language and running machine.
In this framework, the ML process (usually runs on Graphics
Processing Unit (GPU), written in Python) and the CFD process (usually
runs on Central Processing Unit (CPU), written in Fortran/C++)
implemented in different programming languages can work on different
devices to leverage the concise and non-local property of CNN while
maintaining efficiency.

The rest of the paper is organized as follows. In $\S$
\ref{sec:numerical}, we will introduce the numerical methods of ML
and CFD, as well as the details of the HML-CFD framework. The results
of $priori$ test and $posteriori$ test in turbulent channel flow are
discussed in $\S$ \ref{sec:$priori$} and $\S$ \ref{sec:$posteriori$},
following the investigation of numerical stability and efficiency in
$\S$ \ref{sec:resolution}. Finally, conclusions are
addressed.

\section{Numerical methods}\label{sec:numerical}

The purpose of this work is to establish a mapping from the filtered
velocity field ($\overline{u}$, $\overline{v}$, $\overline{w}$) to SGS
stress $\tau_{ij}$ using a CNN-based model. In this paper, the
datasets used for training and testing ML models come from the DNS
of turbulent channel flow. There are various flow characters in such
wall-bounded flow, which increases the difficulty of modelling.
Besides, the progress of ML-based SGS models in channel flow is
slow and demand for more research.

\subsection {Details of DNS}

\begin{figure}
  \centerline{\includegraphics[width=0.9\textwidth]{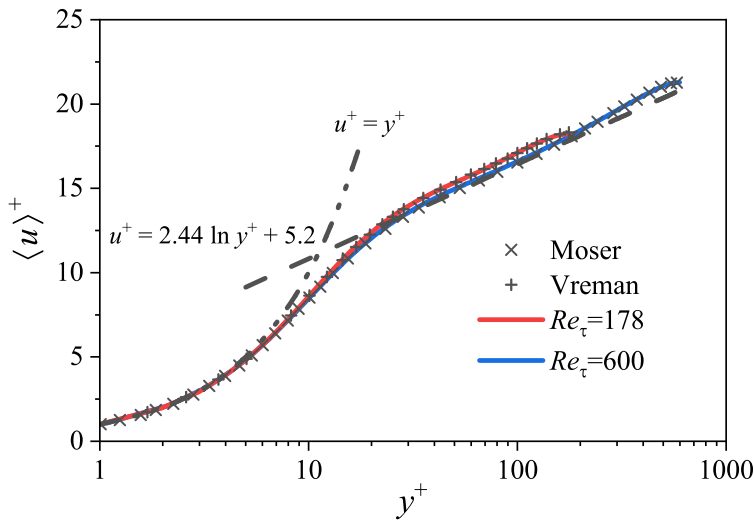}}
\caption{Mean streamwise velocity profiles in wall unit at
$Re_{\tau}=178$, $Re_{\tau}=600$ and corresponding reference data
from \cite{moser1999direct} and \cite{vreman2014comparison}.}
\label{fig:dns}
\end{figure}

The governing equations for DNS of turbulent channel flow are given
by \begin{equation}\label{equ:ns1} \frac{\partial
\mathbf{u}}{\partial t}+\mathbf{u} \cdot \nabla \mathbf{u}=-\nabla
p+\nu \nabla^{2} \mathbf{u}, \end{equation}
  \begin{equation}\label{equ:ns2}
  \nabla \cdot \mathbf{u}=0,
               \end{equation}
where $\mathbf{u}$ is the velocity vector, $p$ is the pressure and
$\nu$ is the kinematic viscosity. The simulation is performed with
periodic boundary conditions in the longitudinal ($x$) and transverse
($z$) directions, the no-slip boundary condition is employed at the
top and bottom walls. A third-order Runge-Kutta and six-order
compact schemes are applied for time integration and spatial
derivatives, respectively. A constant mass flux is maintained in the
channel. Simulations are performed at $Re_{\tau}=178$ and
$600$ with an open-source flow solver Xcompact3d
\citep{bartholomew2020xcompact3d}. Here, $Re_{\tau}$ is the friction
Reynolds number defined by wall fraction velocity $u_{\tau}$, the
kinematic viscosity $\nu$ and the channel half-with $\delta$, while
the bulk Reynolds number is denoted as $Re_{b}$. The computational
domain sizes of DNS in the three dimensions are
$L_x=4\pi$, $L_y=2$, and $L_z=2\pi$. The other computational parameters are
listed in table \ref{tab:dns}. The validity of the DNS is shown in
figure \ref{fig:dns}. To obtain the filtered data from DNS result,
the sharp spectral filter is used in the wall-parallel ($x$ and $z$)
directions as in the preview study \citep{park2021}. The filter
kernel of sharp spectral filter in the spectral space is
$\hat{G}(\boldsymbol{k}) = H(k_{c,x}-|k|)H(k_{c,z}-|k|)$, where $H$
is the Heaviside function. The cut-off wavenumbers in the $x$ and $z$
directions are $k_{c,x}=24(2\pi/L_x)$ and $k_{c,z}=24(2\pi/L_z)$
respectively, which correspond to the filter size $(\overline{\Delta
x}^{_+}$, $\overline{\Delta z}^{_+}$) = (46.7, 23.4) in $Re_{\tau}=178$
case. The training dataset is sampled every 8 and 4 grid points in the
$x$ and $z$ direction respectively, so that CNN can be
trained with the similar grid resolution as the actual LES. All the
grid points in the $y$ direction are sampled except for that on the
wall. The grid sizes used in the DNS of $Re_{\tau}=178$ case are $N_x, N_y,
N_z=384,129,192$ in the three directions. Using the above mentioned
sampling method, totally 292608 ($N_x/8\times (N_y-2)\times N_z/4$)
grid points are sampled at each instant. We collect 200 instantaneous
fDNS fields from DNS data at $Re_{\tau}=178$ ($ 80\% $ for training,
$10\%$ for validation and $10\%$ for the $priori$ test in $\S$
\ref{sec:$priori$}). Based on our test, we also identify that
no noticeable improvement of the performance reaches if using more training data.

\begin{table}
  \begin{center}
\def~{\hphantom{0}}
  \begin{tabular}{lccccc}
  $Re_b$ & $Re_{\tau}$ & $N_x, N_y, N_z$ & $L_x, L_y, L_z$ & $\Delta x^{+}, \Delta y_{min}^{+}, \Delta z^{+}$ \\
  4200 & 178 & 384, 129, 192 & $4\pi, 2, 2\pi$ & 4.4, 1.0, 4.4 \\
  16800 & 600 & 384, 321, 384 & $2\pi, 2, \pi$ & 9.8, 1.0, 4.9 \\
  \end{tabular}
\caption{Parameter values of DNS. Here, $L_x, L_y, L_z$ are the
computational domain sizes in the streamwise (x), wall-norm ($y$) and
spanwise ($z$) direction, respectively. $N_x, N_y, N_z$ are the
corresponding grid sizes. $\Delta x^{+}$ and $\Delta z^{+}$ denote
the grid resolution in wall units, $\Delta y_{min}^{+}$ is the
finest resolution in the wall-normal direction.}
  \label{tab:dns}
  \end{center}
\end{table}

\subsection {CNN-based SGS stress model}\label{sec:cnnbased}

Artificial neural network and convolutional neural network are the
most widely used tools in the machine learning field. As shown in
figure \ref{fig:ann_cnn}(a, b), there are multiple layers
between the input and output layers for both networks, each layer's
input comes from the output of its former layer. In ANN, each neuron
in one layer is usually connected to all neurons in the layer before
it, the procedure of computing layer output can be formalized as
\begin{equation}\label{equ:NN}
  X^l_i = \phi (\sum_{j=1}^{f_n^{'}} W^{l}_{i,j} X^{l-1}_{j}+b^{l}_i),\end{equation}
where $X_{i}^l$ denotes the $i$th output in layer $l$, $\phi$ is the
nonlinear activation function, $f_n^{'}$ is the number of hidden neurons in layer $l-1$, $W^l$ and $b^l$ are the trainable
parameters called weights and bias, respectively.

For CNN, each neuron only receives a restricted region of data from
the previous layer called the neuron's receptive field. Trainable
weights are called filters in CNN, which are shared across neurons
in the same layer and convolves on receptive fields in an iterative
way to get the feature maps. The convolution operation is expressed
as
  \begin{equation}\label{equ:cnn}
  X^{l}_{i,j,k} = \phi(\sum_{m=- f_h/2}^{f_h/2}\sum_{n=- f_w/2}^{f_w/2}\sum_{k^{'}=1}^{f_n^{'}} W^{l}_{m,n,k^{'},k}X^{l-1}_{i+m,j+n,k^{'}} + b^{l}_k ),
\end{equation}
where $f_h$ and $f_w$ denotes height and width of
filters, respectively. $f_n^{'}$ is the depth of filters in the previous layer (layer $l-1$), $X^{l}_{i,j,k}$ is the output of the neuron
located in row $i$, column $j$ in the feature map $k$ of the convolution
layer $l$. $W^{l}_{m,n,k^{'},k}$ is the connect weight between
feature map $k^{'}$ (layer $l-1$) and feature map $k$ (layer $l$),
while $b^{l}_k$ denotes the corresponding bias term. Weight
sharing makes the extraction of shift-invariance and hierarchical
features possible in deep neural networks. Because of the distinct
character, CNN is ideal for data with a grid-topology such as image
and CFD data.

As shown in figure \ref{fig:ann_cnn}(b), the prediction on a point in feature map $X^{l+1}$ totally relies on its receptive field (e.g. $3\times3$) on the feature map $X^{l}$, while the prediction of a point in $X^{l}$ depends on its receptive fields (e.g. $3\times3$) on $X^{l-1}$, which means a $5\times5$ size receptive fields in layer $l-1$ contributes to the prediction of the point in layer $l+1$. With the deepen of the neural network, the receptive field gets larger. Generally, the receptive field of a CNN with kernel size $f_h\times f_w$ and $L$ hidden layers is $[L(f_h-1)+1] \times [L(f_w-1)+1]$. In other word, information from a $[L(f_h-1)+1] \times [L(f_w-1)+1]$ size regime around a point directly or indirectly influences the final prediction on the point. Therefore, we could use CNN to develop a nonlocal modelling method.

In terms of the SGS stress modelling problem, a CNN-based model can
directly handle the raw field variables
$\overline{u}$, $\overline{v}$, and $\overline{w}$ without destroying the
spatial topology and the flow field feature extraction is automatic. However, the feature selection is hand-engineered and depends on $priori$ knowledge for ANN. On the other hand,
the ANN-based model can only include spatial
information by adding more neighbouring gird points into input
features, which multiplies the number of input features and enlarges
the model size.

The training data are the same for ANN and CNN but organized in
different ways. ANN makes predictions in a point-to-point way, i.e.
taking physical values on a grid point (data shape $N_{feature}$)
and predicts $\tau_{ij}$ on this grid point, while CNN works in a
block-to-block way. Here, $N_{feature}$ is the number of input
features. CNN can take flow variables on the $x$-$z$ plane (data shape
$N^{LES}_x \times N^{LES}_z \times N_{feature}$ corresponding to the
two-dimensional (2D) convolution kernel) or a whole three-dimensional
(3D) flow field (data shape $N^{LES}_x \times N^{LES}_y \times N^{LES}_z
\times N_{feature}$ corresponding to the 3D convolution
kernel) as the input. The grid of turbulent channel flow is usually
uniform in the $x$ and $z$ direction but nonuniform in the $y$ direction,
the traditional 3D convolution kernel cannot represent uniform grid
spacings in different directions. Besides, the 2D convolution kernel
requires fewer parameters, therefore it is more efficient than the
3D one. Therefore, we choose the $x$-$z$ plane type input in this
work. Specifically, the training data are fed into the CNN in the shape of $48\times48\times3$ ($=N_x/8\times N_z/4 \times N_{feature}$). The information of the $y$ direction is contained in the training data sampled at different $y$ locations. This kind of input can be
applied to different grid resolutions in the $y$ direction, thus more
flexible.

\begin{figure}
  \centerline{\includegraphics[width=0.9\textwidth]{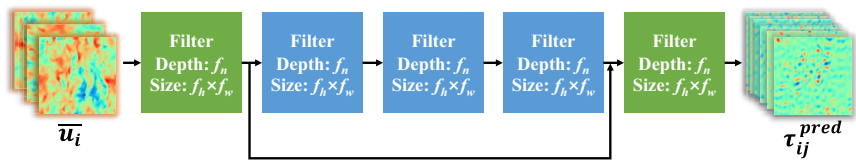}}
\caption{Schematic of the ResNet. The inputs and outputs are the
primitive quantities $\bar{u}$, $\bar{v}$, and $\bar{w}$ and six components of
$\tau_{ij}$, respectively. For each convolutional layer, the depth
of feature maps is $f_n$ and the filter kernel size is $f_h \times f_w$.}
\label{fig:resnet}
\end{figure}

The CNN used in this work is a ResNet \citep{he2016deep} with 5
hidden layers shown in figure \ref{fig:resnet}. The input is the
three components of filtered velocity ($N_{feature}=3$), while the
output is the corresponding six components of SGS stress
$\tau_{ij}$. The filter kernels in each hidden layer consist of $f_{n}$
feature maps with size $f_h\times f_w$. Here $f_h$ and $f_w$ represent the sizes of receptive fields in each layer which can be regarded as a measure of non-local effect. Generally, the square filter ($f_h = f_w$) is the most widely used in practice, which is also adopted in this work. A series of CNN-based models with different filter depth and  size are examined in $\S$ \ref{sec:$priori$}. As a comparison, an ANN with five hidden layers ($f_{n}$ neurons per hidden layer) is also
employed. Note that the input feature selection is crucial for
ANN-based models, but it is not the focus of this article. The
velocity gradient tensor $\partial{u_{i}}/\partial{x_j}$ is chosen
as the input feature of ANN ($N_{feature}=9$), since it is widely
used in preview works \citep{gamahara2017, wang2018, park2021}, while
the output is also $\tau_{ij}$. The activation function $\phi$ for
both models is the exponential linear unit (ELU) defined as
\begin{equation} f(x)=\left\{\begin{array}{ll}
x & , if \quad x>0 \\
\alpha(e^x-1) & , if \quad x \leq 0
\end{array}\right.,
\end{equation} where $\alpha = 1.0$ here. There is no activation function in the last layer of both ANN and CNN.

For both the ML models, the objective function to minimize is the mean
square error (MSE) loss function denoted as \begin{equation}
l_{MSE}=\frac{\sum_{n=1}^{N}\left\| \tau^{fDNS}_{ij,n}-
\tau^{pred}_{ij,n}  \right\|_{2}^{2}}{N},
\end{equation}
where $\tau^{pred}_{ij}$ is the output of ML models,
$\tau^{fDNS}_{ij}$ is computed from fDNS data using equation
$\tau^{fDNS}_{i j}=\overline{u_{i} u}_{j}-\bar{u}_{i} \bar{u}_{j}$
and $N$ is the total number of training data. The Adam optimizer \citep{kingma2014adam} is
employed to minimize the loss function. We use the friction velocity
$u_{\tau}$ to norm the training data.

\subsection {HML-CFD framework}\label{sec:hmlcfd}

\begin{figure}
  \centerline{\includegraphics[width=0.9\textwidth]{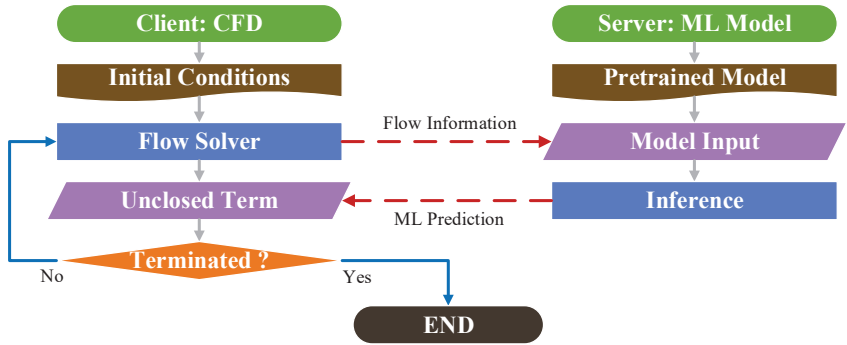}}
  \caption{Flow chart for the heterogeneous ML-CFD (HML-CFD) framework.}
\label{fig:hml_cfd}
\end{figure}

It is known that the ML algorithms (e.g. ANN and CNN) written mostly
in Python are suitable for devices with a highly parallel structure
like GPU, while most CFD algorithms written in Fortran/C++ run on CPU.
When applying ML models to realistic numerical simulations, we may
meet the problem of devices selection and mix-language programming.
The HML-CFD framework is designed to make the combination of ML
models and CFD simulations concise and efficient. As depicted in
figure \ref{fig:hml_cfd}, the framework consists of a client and a
server. The client is the CFD simulation process demanding unclosed
terms or undetermined coefficients, while the server is the ML
process waiting for the request from the client. The client sends
fluid information to the server first and then waiting for the
prediction of pre-trained ML models, the CFD process goes on after
the predicted result is returned. The two processes exchange
messages through an inter-process communication technology called
sockets. A socket is one endpoint of a two-way communication link
between two programs running on the network. Sockets allow
communications between two different processes on the same or
different machines. With the HML-CFD framework, we can focus on the
development of the CFD process, while the ML process can work on
efficient parallel devices like GPU or cloud computing resources.

In the case of CNN-based SGS model, the CFD process firstly sends
the fluid field data $\overline{u}$, $\overline{v}$, and $\overline{w}$ to the
ML process, then ML model predicts the SGS stress $\tau_{ij}$
according to the received messages and returns the predictions to the
CFD process. This procedure repeats until the end of CFD
simulation. Owing to the independence of the two program processes,
the CFD program and ML program can run on different machines using
different languages. In this way, we can take advantages of parallel
computing in the computation of ML models and avoid the trouble of
mixed-language programming. Hence, our proposed technique is
applicable to the actual LES efficiently.

In this work, the CFD simulations are implemented by Fortran on
Intel Core i7-9700K (CPU), while the ML algorithms are conducted
using the open-source library Pytorch \citep{paszke2019pytorch} on
NVIDIA GeForce GTX 2080 (GPU).

\section {Results}\label{sec:result}

To examine the performance of ML-based models, a $priori$ and a
$posteriori$ test are conducted in $\S$ \ref{sec:$priori$} and $\S$ \ref{sec:$posteriori$},
respectively. In the $priori$ test, the SGS stresses are predicted by
ML models (ANN and CNN) with the input variables from fDNS at
$Re_{\tau}=178$. Two traditional models, i.e. Smagorinsky (SM) model, and Wall-Adapting Local Eddy-Viscosity (WALE) model are also performed as comparison. The WALE model is a kind
of algebraic eddy viscosity model, it can return the correct
wall-asymptotic behaviour, thus suitable for wall-bounded flows like
channel flow. In the $posteriori$ test, all the models are applied in
actual LESs. Finally, the numerical stability and computational
efficiency are discussed in $\S$ \ref{sec:resolution}.

\subsection{A $priori$ test}\label{sec:$priori$}

The pre-trained models are examined with test data from fDNS at
$Re_{\tau}=178$ firstly. The correlation coefficients between the models' prediction $\tau^{pred}_{ij}$ and $\tau^{fDNS}_{ij}$, i.e.
 \begin{equation}\label{equ:rho}
  \rho_{\tau}= \frac{\langle(\tau^{pred}_{ij}-\langle \tau^{pred}_{ij}\rangle)(\tau^{fDNS}_{ij} - \langle \tau^{fDNS}_{ij} \rangle) \rangle}{\sqrt{\langle (\tau^{pred}_{ij}-\langle\tau^{pred}_{ij}\rangle)^2 \rangle}\sqrt{\langle (\tau^{fDNS}_{ij}-\langle\tau^{fDNS
  }_{ij}\rangle)^2 \rangle}},
 \end{equation}
is used to assess models' performance. Totally 4 CNN-based models, i.e. CNN-K1, CNN-K3, CNN-K5, CNN-K7, are examined here. CNN-K3 denotes the size of filters used in this network is $f_h\times f_w=3\times3$, CNN-K5 and CNN-K7 have the similar meanings. CNN-K1 is equivalent to an ANN with raw quantities $\overline{u}$, $\overline{v}$, and $\overline{w}$ as input, since the filter size is $f_h\times f_w=1\times1$, which means nonlocal information is not considered.

\begin{figure}
  \centering
  {\includegraphics[width=0.475\textwidth]{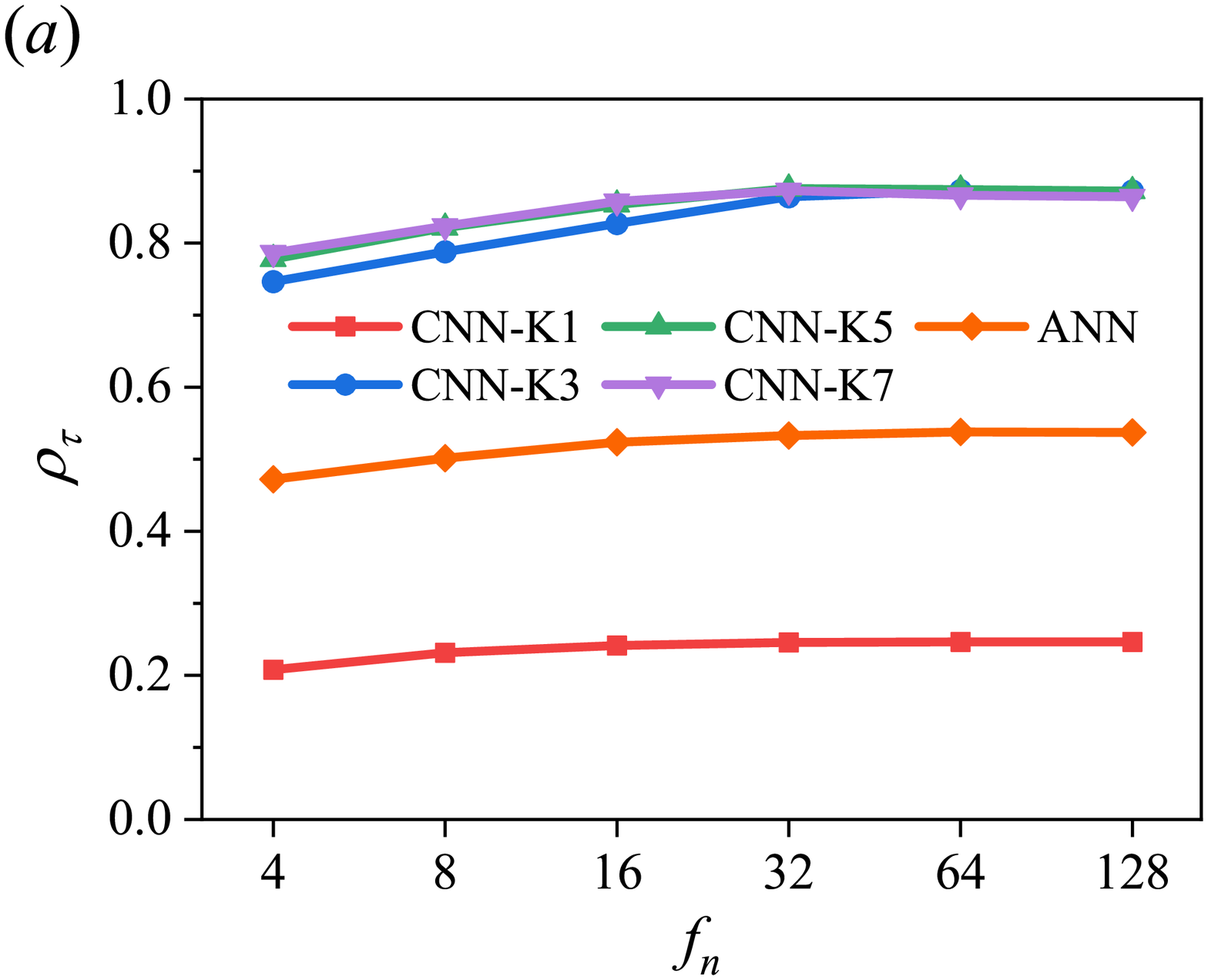}}
  {\includegraphics[width=0.475\textwidth]{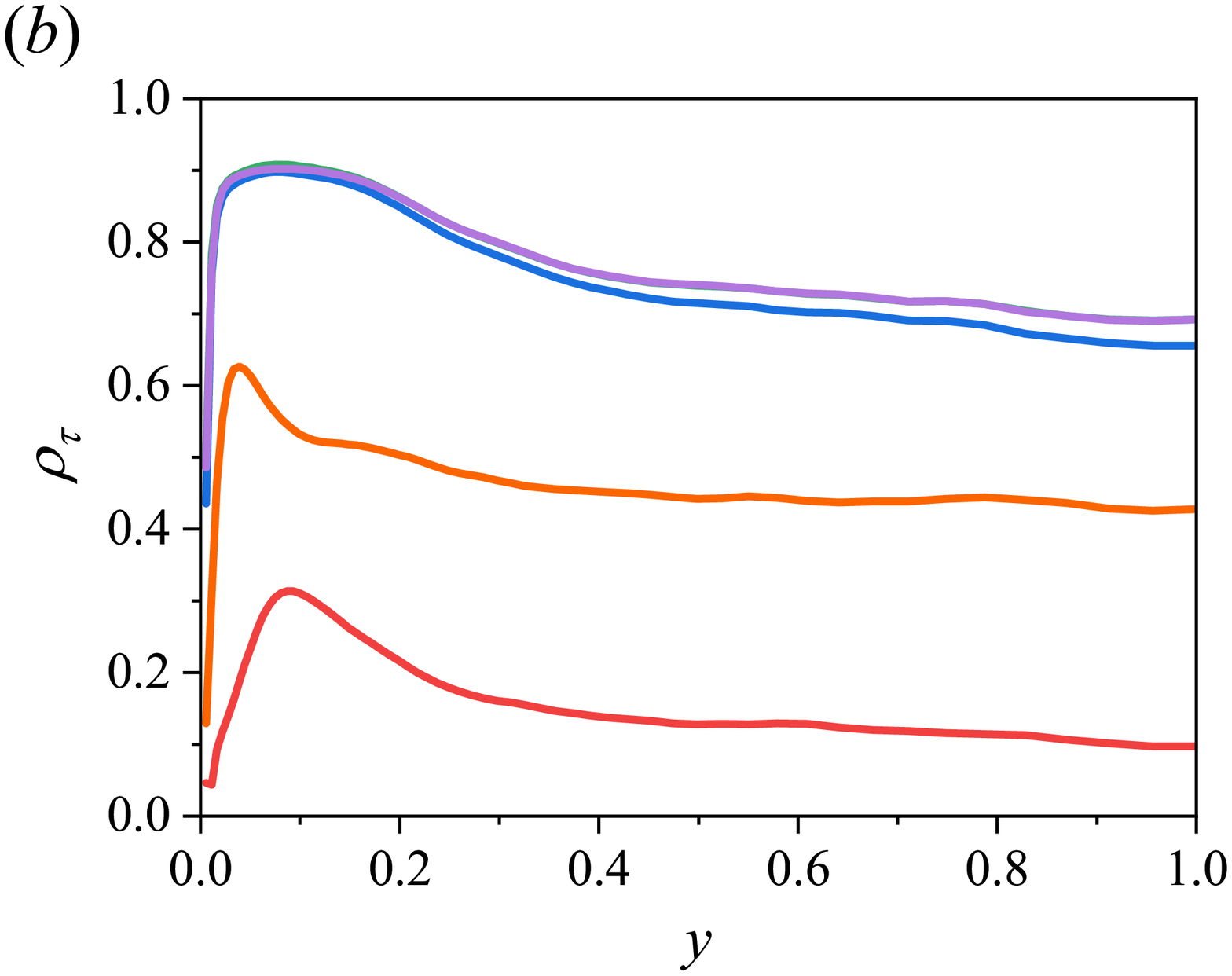}}
\caption{Correlation coefficients between the true and predicted
$\tau_{ij}$ from different models. Correlation coefficients are
averaged over the whole domain (a) and $x$-$z$ plane (b),
respectively.} \label{fig:rho}
\end{figure}

To identify the relatively optimal parameters for different data-driven models, the effect of filter depth $f_n$ (or hidden neurons numbers per layer for ANN) is firstly investigated. As presented in figure \ref{fig:rho}(a), with the increase of $f_n$, $\rho_{\tau}$ slightly raises till a critical $f_n$, after which $\rho_{\tau}$ shows no noticeable change, indicating larger $f_n$ is unhelpful. ANN and CNN-K1 are local methods, they only have different input features, i.e. $\partial \overline{u}_i /\partial x_j$ and $\overline{u}_i$, respectively. $\rho_{\tau}$ of ANN (about $0.5$) is much larger than that of CNN-K1 (about $0.24$), exhibiting the importance of feature selection in the local method.  For the nonlocal CNN-based model with primitive flow variables $\overline{u}$, $\overline{v}$, and $\overline{w}$ as input, i.e. CNN-K3, CNN-K5, and CNN-K7, their predictions show relatively high correlation ($\rho_{\tau}$ above $0.8$) with the reference data and significantly higher than that of ANN and CNN-K1, indicating the CNN successfully extracted more effective features than the velocity gradient tensor related to the SGS stress from the primitive flow variables. From the comparison of the nonlocal CNN-based models, we could find the larger kernel size, including more nonlocal information in prediction, generally leads to better results but the tread is not so obvious after $f_h$ and $f_w$ greater than $3$. The performance of CNN-K5 and CNN-K7 is nearly the same (marginally outperforms CNN-K3), which demonstrates moderately introducing nonlocal information is beneficial for the LES modelling.

Considering the balance of accuracy of efficiency, CNN-K3, CNN-K5, and CNN-K7 with $f_n=32$, CNN-K1 and ANN with $f_n=16$ are selected as the tested models in the following work. The variation of $\rho_{\tau}$ with $y$ is depicted in figure \ref{fig:rho}(b). It is seen that $\rho_{\tau}$ of all models is relatively uniform in the outer layer but varies dramatically in the inner layer. The nonuniform performance may result from the fact that there exist distinct flow characters in different flow regions, while here only one model is used to represent features in the whole flow regime. The $\rho_{\tau}$ from nonlocal data-driven models keep in a relatively high level in all the wall-normal distances especially for the near-wall vicinity. The results of CNN-K5 and CNN-K7 are quite close, slightly larger than that of CNN-K3 in the outer layer, which is consistent with the result displayed in figure \ref{fig:rho}(a).

Therefore, the CNN-based models successfully extracted the key flow features from primitive variables, the involved nonlocal information contributes to the excellent performance in the $priori$ test. However, high correlation coefficients of $\rho_{\tau}$ in $priori$ test can not guarantee the success in the actual LES \citep{park2021}, it is necessary to access the accuracy, stability and efficiency of SGS stress models in $posteriori$ test.

\subsection{A $posteriori$ test}\label{sec:$posteriori$}

LES of turbulent channel flow with a constant mass
flow at $Re_b=4200$ (denoted as LES178) and $Re_b=16800$ (denoted as LES600) will be performed using
different models. All the LESs are carried out with the same numerical
methods as those of DNS described in $\S$ \ref{sec:cnnbased}. The
computational parameters of LESs are listed in table \ref{tab:les}.
The time steps of LESs are 8 times that of DNS. Note that the $van
Driest$ damping function \citep{moin1982numerical} (multiplying the SGS stress by $(1-e^{-y^{+}/A^{+}})^2$ with $A^{+}=25$) is used in the
$posteriori$ test of the SM model, no extra treatment is needed in
the other models. $Re_{\tau}$ obtained from LES with nonlocal models are close to
that of DNS (less than 3 \% error) in LES178 case. ANN with $\partial \overline{u}_i/\partial x_j$ as the input shows similar performance as traditional models (around 5\% error), while CNN-K1 underpredicts $Re_{\tau}$, indicating the friction velocity $u_{\tau}$ is underestimated seriously.

\begin{table}
  \begin{center}
\def~{\hphantom{0}}
  \begin{tabular}{lcccccc}
  Case & $Re_{b}$ & $(N_x, N_y, N_z)$ & $L_x, L_y, L_z$ & $(\overline{\Delta x}^{_+}, \overline{\Delta z}^{_+})$ & SGS model & $Re_{\tau}$ \\
  LES178 & 4200 & 24, 49, 24 & $2\pi, 2, \pi$ & 46.7, 23.4 & SM & 169 \\
         & --- &   ---       & --- & ---     & WALE & 170\\
         & --- &   ---       & --- & ---    & ANN & 169\\
         & --- &   ---       & --- & ---    & CNN-K1 & 164\\
         & --- &   ---       & --- & ---    & CNN-K3 & 175\\
         & --- &   ---       & --- & ---    & CNN-K5 & 173\\
         & --- &   ---       & --- & ---    & CNN-K7 & 180\\
  LES600 &16800&  42, 97, 42 & $\pi, 2, \pi/2$ & 44.8, 22.4 & SM & 571\\
         & --- &   ---       & --- & ---    & WALE & 570\\
         & --- &   ---       & --- & ---    & ANN & diverged\\
         & --- &   ---       & --- & ---    & CNN-K1 & 563\\
         & --- &   ---       & --- & ---    & CNN-K3 & 594\\
         & --- &   ---       & --- & ---    & CNN-K5 & 595\\
         & --- &   ---       & --- & ---    & CNN-K7 & 595\\
  \end{tabular}
\caption{Parameter values of LES, $\overline{\Delta x}^{+}$ and
$\overline{\Delta z}^{+}$ are computed with $u_{\tau}$ from DNS
(table\ref{tab:dns}).}
  \label{tab:les}
  \end{center}
\end{table}

\begin{figure}
  \centering
  {\includegraphics[width=0.475\textwidth]{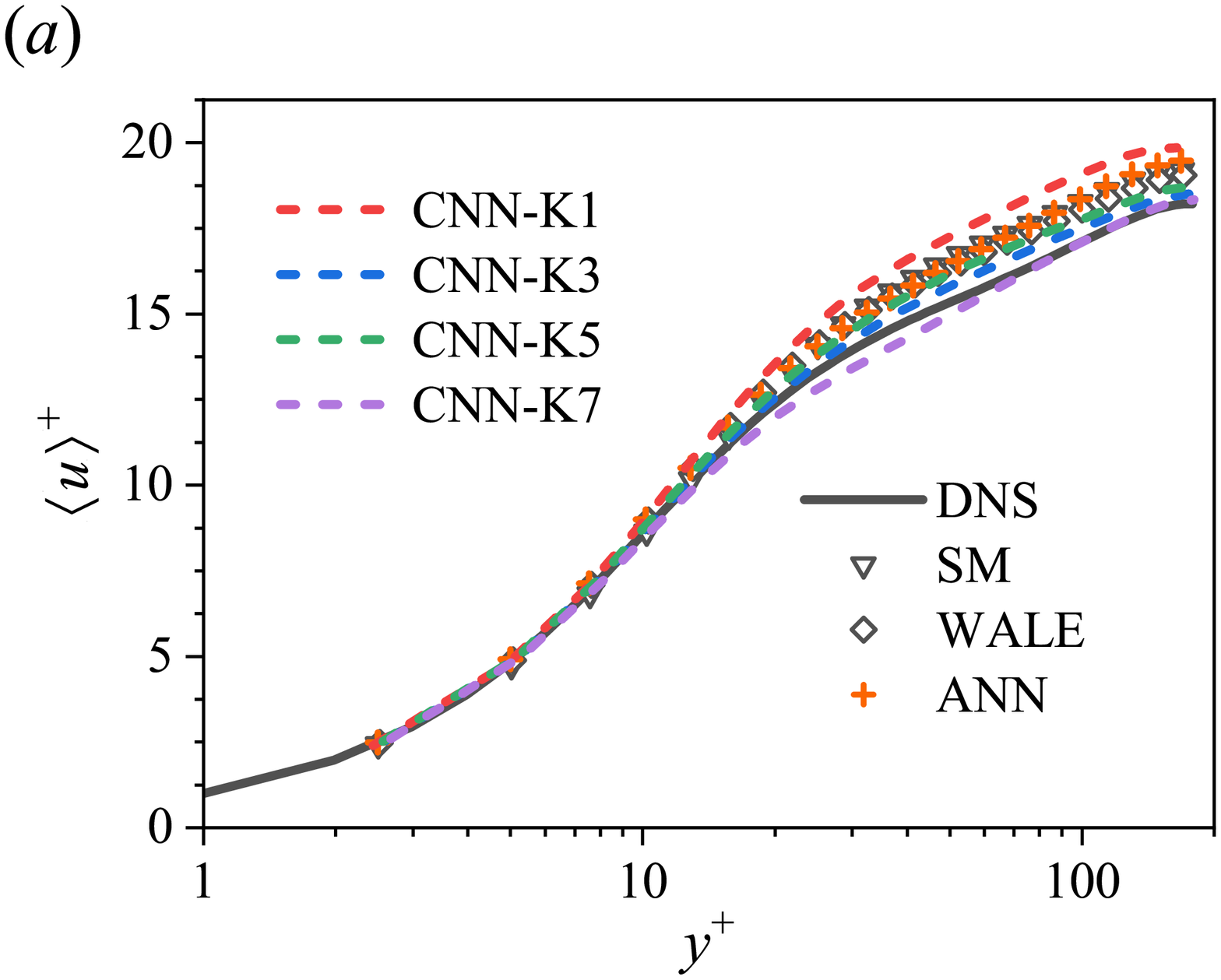}}
  {\includegraphics[width=0.475\textwidth]{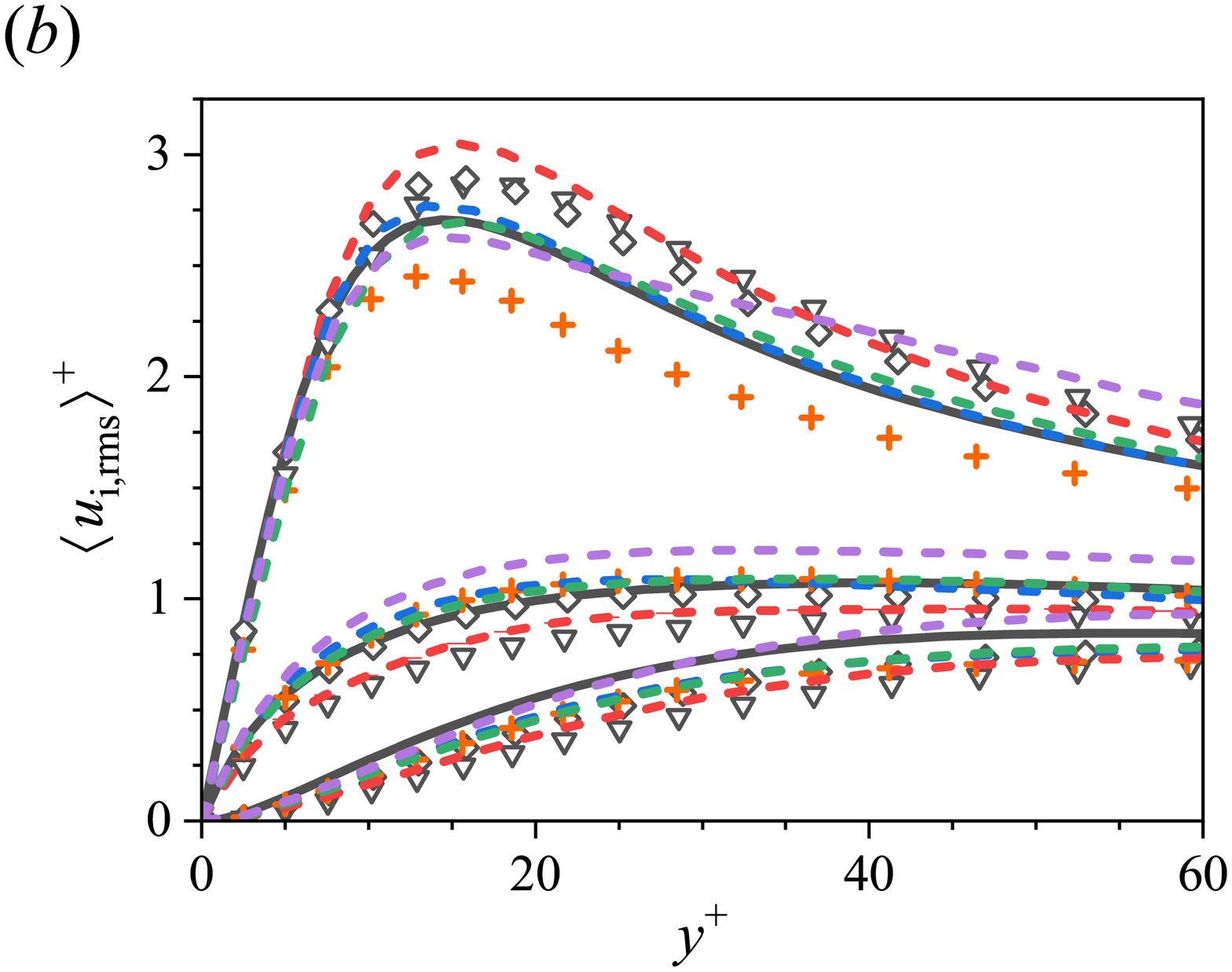}}
\caption{Comparison of mean flow (a) and root-mean-square (r.m.s.) of velocity
fluctuations (b) obtained by LESs with different models at $Re_b =
4200$.} \label{fig:postu}
\end{figure}

Figure \ref{fig:postu}(a) compares the mean velocity profiles from LES simulations at $Re_b = 4200$. CNN-K3 and CNN-K5 show excellent predictions of the mean velocity, while local methods and the two traditional models overestimate the profile because of the underestimated $u_{\tau}$. Among all the models, only CNN-K7 predicts a lower mean velocity profile. In terms of velocity fluctuations, CNN-K3 and CNN-K5 also perform good, providing fairly well agreements of result with the reference in the $x$ and $z$ direction but noticeably deviation in the $y$ direction. It is difficult to capture fluctuations in the $y$ direction since the value is too small compared to the streamwise fluctuations, especially in near-wall regions. Note that CNN-K7 does not well predict the velocity fluctuations, the root-mean-square (r.m.s.) in the $x$ direction decays too slow with the increase of the wall-normal distances after the peak value. CNN-K3 surpasses CNN-K7, even $\rho_{\tau}$ of CNN-K3 is lower in the prior test. On the other hand, the ANN-based model generates fewer fluctuations than the DNS result while CNN-K1 and traditional models provide more. In general, all the SGS models generate reasonable solutions and the nonlocal data-driven models (except for CNN-K7) shows advantages over traditional models, while local ML models (i.e. CNN-K1 and ANN) do not.

\begin{figure}
  \centering
  {\includegraphics[width=0.475\textwidth]{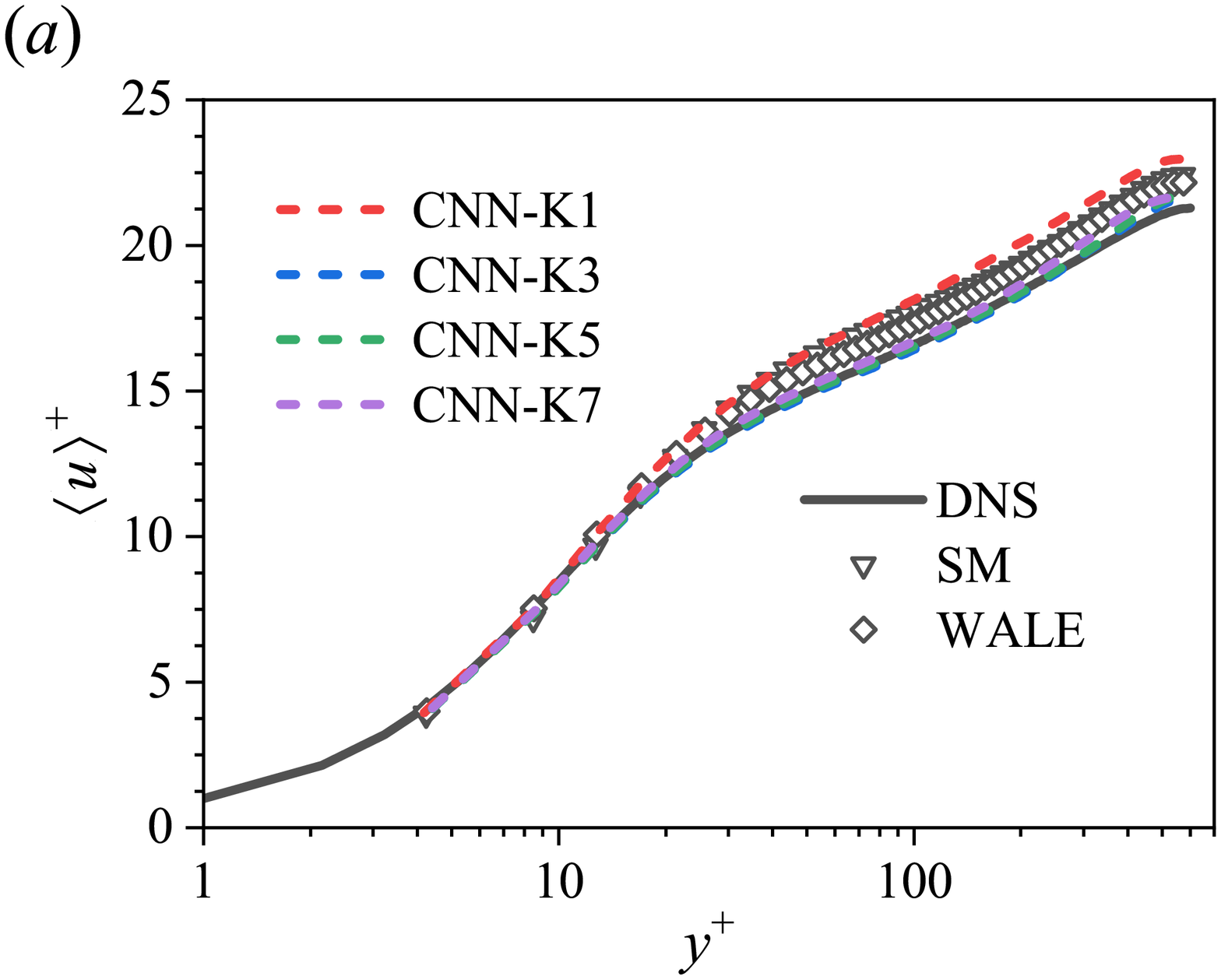}}
  {\includegraphics[width=0.475\textwidth]{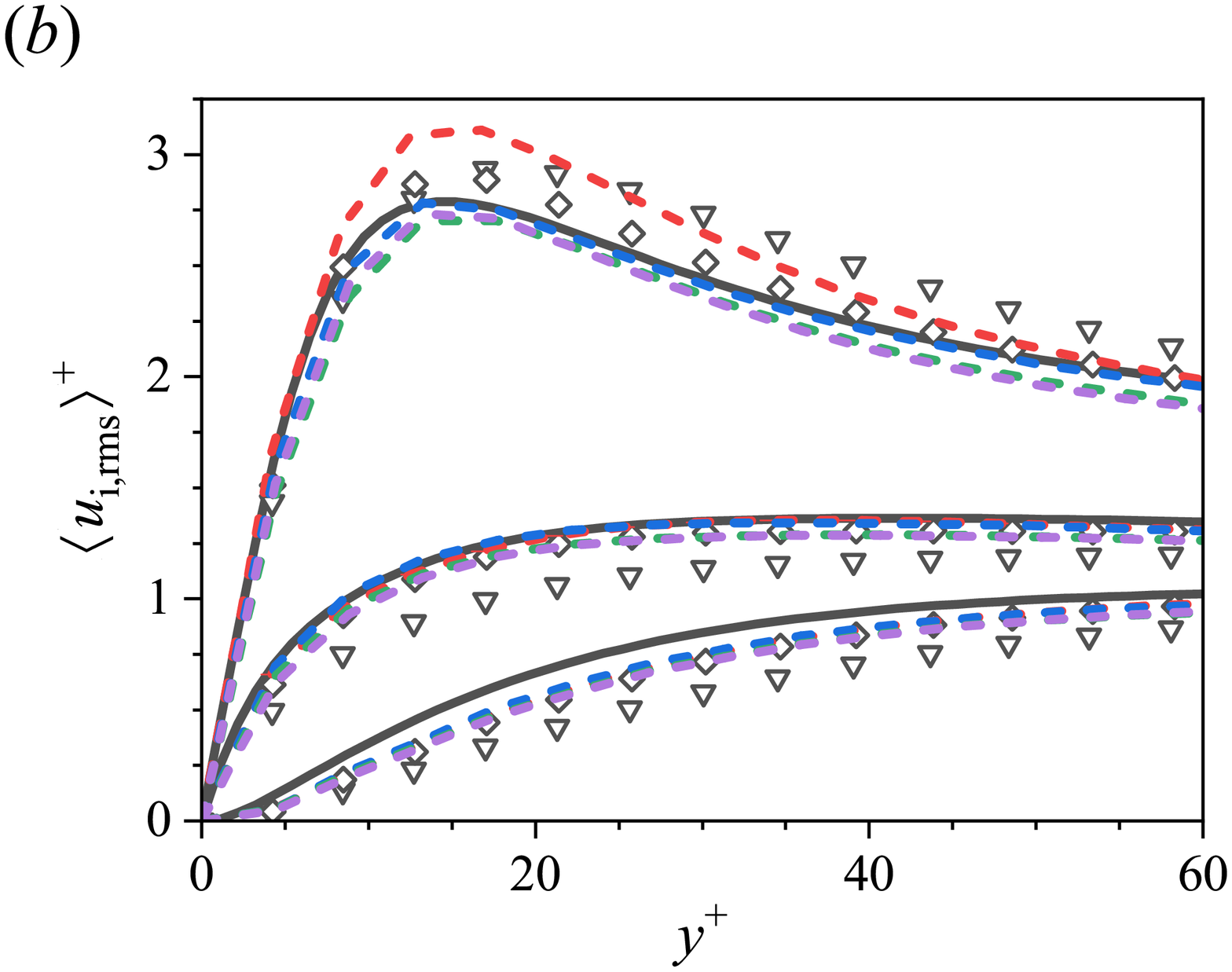}}
\caption{Comparison of mean flow (a) and r.m.s. of velocity
fluctuations (b) obtained by LESs with different models at
$Re_b=16800$.} \label{fig:postu600}
\end{figure}

The previous test is conducted at the same Reynolds number ($Re_b = 4200$) as that of training data. To evaluate the extrapolation ability of ML models, the $posteriori$ test is also carried out at a higher Reynolds number $Re_b = 16800$. The ANN-based model diverges in this case, numerical stability is a big issue for the local data-driven method like the ANN-based model. As demonstrated in \cite{park2021}, most ANN-based models face the numerical divergence problem without special treatments (e.g. clipping backscatter), while the extra treatments introduce human intervention and break the intention of data-driven methods to some extend. The SM and WALE models are numerically stable in a wide range of Reynolds numbers, which is the advantage of traditional models with no backscatter. All the CNN-based models generate physically reasonable solutions just as their performance in LES178 case. $Re_{\tau}$ from LES600 are well predicted by the nonlocal models (less than 2\% error), while CNN-K1 and the two traditional models underpredict it. Figure \ref{fig:postu600} (a, b) shows the CNN-K3 well capture the mean velocity profile and the r.m.s of velocity fluctuations, while CNN-K5 and CNN-K7 generate less fluctuation than the reference data but still reasonable. The result of CNN-K1 closes to that of the SM model, both overpredicting the mean flow and $u_{rms}$. LES with the WALE model is better than that of the SM model, the subgrid eddy viscosity $\nu_t$ of the WALE model tends towards zero as approaching the wall, which is more physically reasonable contrast to constant $\nu_t$ in all the wall-normal distances.

\begin{figure}
  \centering
  {\includegraphics[width=0.475\textwidth]{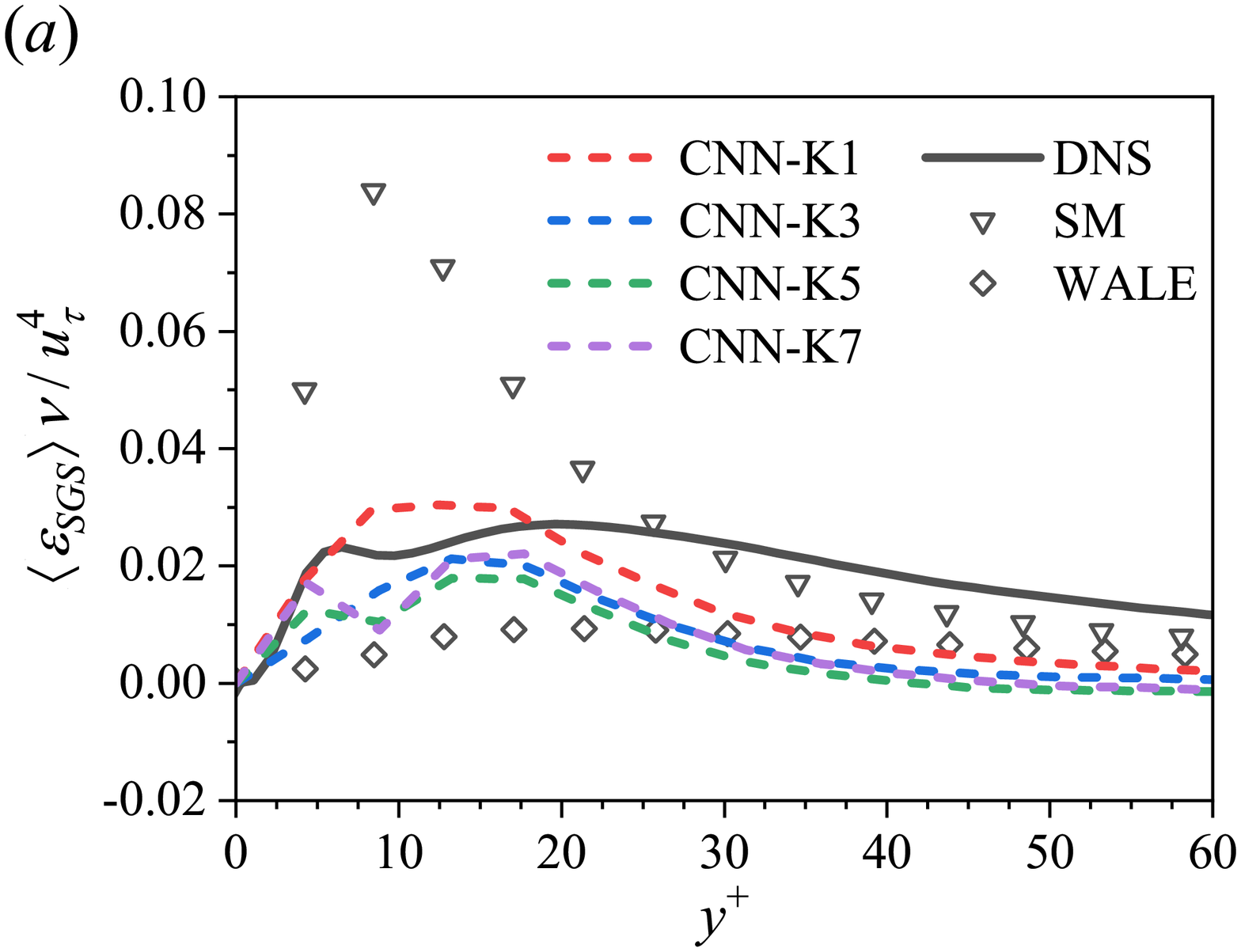}}
  {\includegraphics[width=0.475\textwidth]{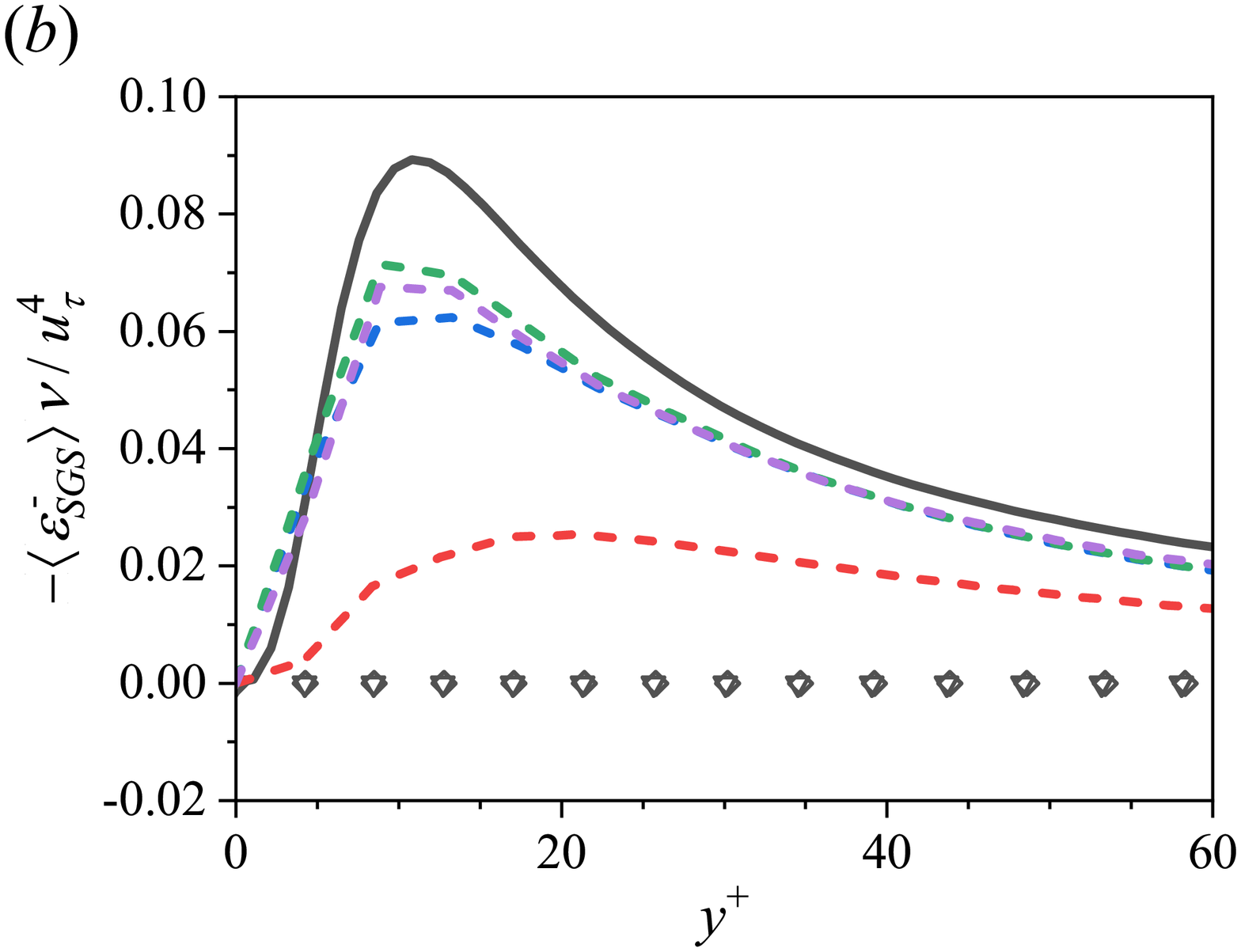}}
  \caption{Turbulent statistics from the $posteriori$ test at $Re_b = 16800$: (a) mean SGS dissipation; (b) mean backscatter.}
\label{fig:postsgs}
\end{figure}

Figure \ref{fig:postsgs} (a, b) compares the prediction of the mean
SGS dissipation ($\epsilon_{SGS} = -\tau_{ij}\overline{S_{ij}}$) and
mean backscatter ($\langle \epsilon^{-}_{SGS} \rangle = \frac{1}{2}
\langle \epsilon_{SGS} - | \epsilon_{SGS} | \rangle$). The SM model
dramatically overestimates the SGS dissipation, which indicates much more
energy is transported from resolved scale to unresolved scale.
On the contrary, the nonlocal CNN-based models slightly underestimate $\epsilon_{SGS}$ in the near-wall regions, while $\epsilon_{SGS}$ predicted by WALE model are too small,
indicating the dissipation is not enough. The backward energy transfer from unresolved scale to resolved scale is the main cause of numerical instability in LES. From figure \ref{fig:postsgs}(b), we could find that the nonlocal models produce slightly less backscatter compared to the reference data without incurring numerical instability, CNN-K5 is the best among CNN-based models. The mean backscatter predicted by CNN-K1 is much lower, which may contribute to its numerical stability in this high Reynolds case. For SM and WALE, only positive $\nu_t$ can be predicted, thus no backscatter is allowed.

\begin{figure}
  \centering
  {\includegraphics[width=0.95\textwidth]{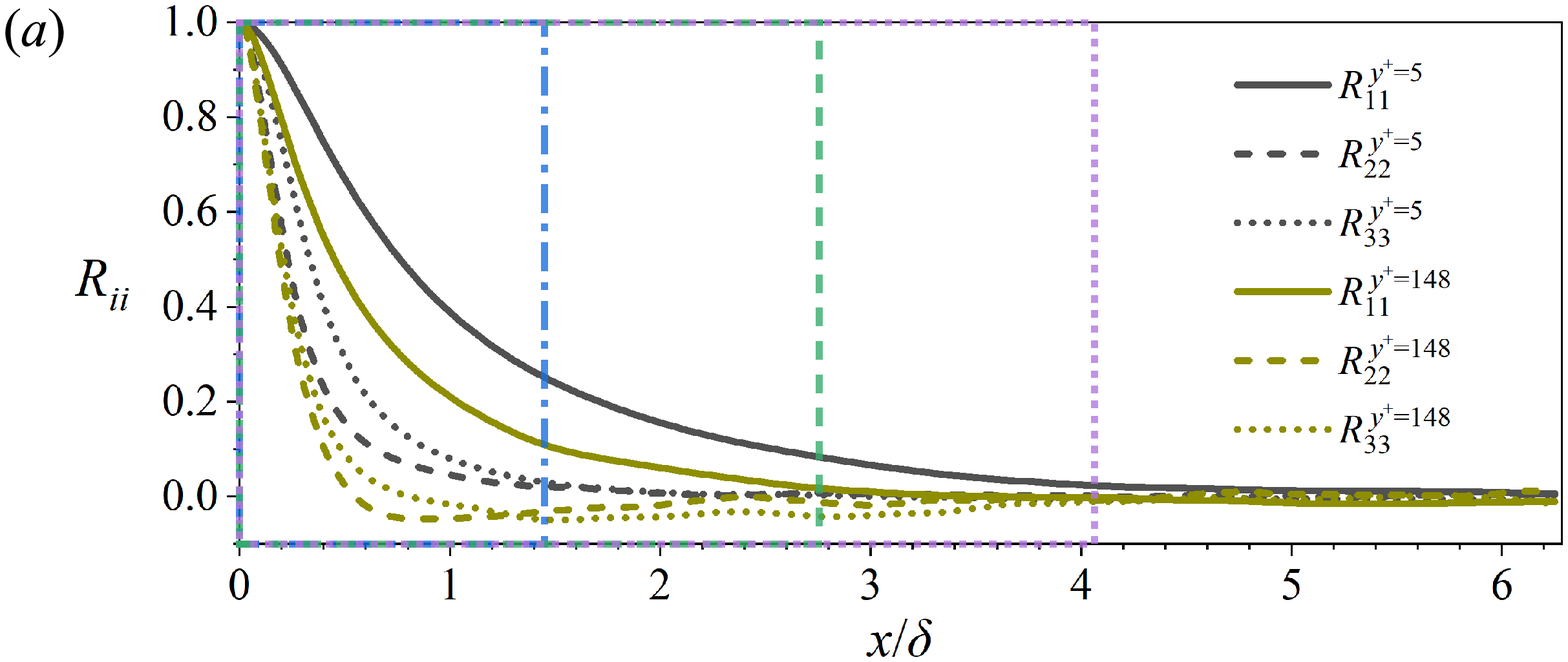}}
  {\includegraphics[width=0.95\textwidth]{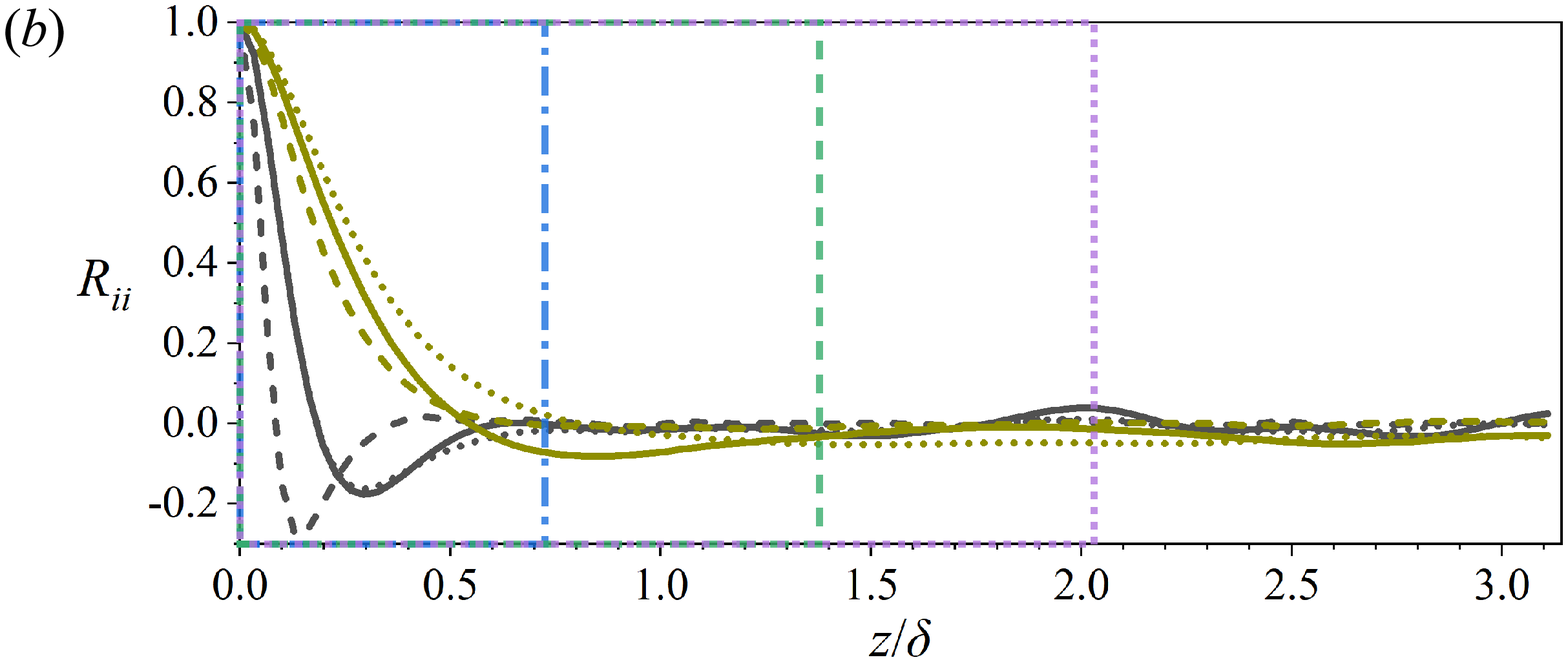}}
  \caption{Two point correlations computed from DNS data ($Re_{\tau}=178$) at $y^+=5$ and $y^+=148$: (a) streamwise separation; (b) spanwise separations. The receptive fields of CNN-based models are represented by rectangles: blue dash dotted line, CNN-K3; green dashed line, CNN-K5; purple dotted line, CNN-K7.}
\label{fig:Rii}
\end{figure}

From previous tests, we could conclude that the nonlocal CNN-based model can well predict the SGS stresses in turbulent channel flow at LES178 and LES600 cases, their performance surpasses both the local data-driven methods (ANN and CNN-K1) and traditional models (SM and WALE model). The extra included spatial information contributes to the success of the SGS stress modelling. However, from the comparison of CNN-K3 and CNN-K7, we could also found that too large receptive fields seem unhelpful for the modelling. To analyze the effect of receptive fields qualitatively and quantitatively, the two-point correlations $R$ in the streamwise and spanwise directions calculated from DNS data at $Re_{\tau}=178$ are presented in figure \ref{fig:Rii}. Two $y^+$ locations are selected, one close to the wall ($y^+=5$) and the other close to the centerline ($y^+=148$). As shown in figure \ref{fig:Rii}, with the separation distance increases, the two-point correlations tend to zero, $R$ fall off steeper in the spanwise direction than that of the streamwise direction, since the streamwise vortices are longer.

In CNN, the final prediction totally depends on the information from its receptive fields, as described in $\S$ \ref{sec:cnnbased}, the receptive field of a CNN with $f_{h}\times f_{w}$ size kernels and $L$ hidden layers is $[L(f_h-1)+1] \times [L(f_{w}-1)+1]$. The receptive field for CNN-K3 is $11\times11$ here. Because the training data is organized in the shape of $48\times48\times3$, the physical domain size corresponding to the receptive field is $\frac{11}{12}\pi \times \frac{11}{24}\pi$ ($4\pi\times \frac{11}{48}=\frac{11}{12}\pi, 2\pi\times \frac{11}{48}=\frac{11}{24}\pi$). In other word, the prediction of $\tau_{ij}$ on a grid point is affected by information from a $\frac{11}{12}\pi \times \frac{11}{24}\pi$ size flow region around it. The receptive field of CNN-K3, as well as that of CNN-K5 and CNN-K7, is indicated by rectangles in figure \ref{fig:Rii}. As shown in figure \ref{fig:Rii}(a), the receptive field of CNN-K3 only contains the region of physical distance less than $\frac{11}{24}\pi$ in the $x$ direction, which includes the most related information to the predicted point in the flow field. For CNN-K5, the scope is larger and includes extra less related information ($R_{ii}<0.3$). CNN-K7 considers nearly $\frac{2}{3}$ of the flow field to make predictions on a grid point, while most of the flow region is irrelative to the grid point, which means much more irrelevant flow field information is included. It increases the difficulty of modelling and risk of overfitting. It is more obvious in the $z$ directions as can be seen in figure \ref{fig:Rii} (b), the two-point correlations $R$ nearly all approach zero when the two-point distance is larger than 0.8 except for $R_{11}$ at $y^+=148$, the extra scopes of CNN-K5 and CNN-K7 do not provide more useful information, which accounts for the reason why CNN-K7 cannot surpass the other nonlocal CNN-based models even with more accessible information. Therefore, introducing moderate nonlocal information is beneficial for the SGS stress modelling, while excessive nonlocal information is unnecessary. The perspective of correlation analysis provides a quantitative method to guide the design of nonlocal data-driven models. On the other hand, the larger kernel size generally leads to more trainable weights and a larger network, thus affecting models' efficiency, which will be discussed below.

\subsection{The numerical stability and computational efficiency}\label{sec:resolution}

From the preceding analysis, the ML-based models are examined in the
aspect of accuracy, while numerical stability and computational
efficiency are also crucial for SGS models. Here, an
extra $posteriori$ test is carried out for the nonlocal CNN-K3 model to
access its numerical stability and investigate the influence of grid resolutions. We select a series of grid resolutions from nearly one half to double the spatial
resolution in the $x$ and $z$ direction used in models' training ($(\overline{\Delta
x}^{_+},\overline{\Delta z}^{_+}) = (46.7, 23.4)$), the simulations are conducted in the channel flow at $Re_{b}=16800$ with domain size $L_x$, $L_y$, $L_z$ = $\pi$, 2, $\pi/2$. The grid numbers $N_x$ and $N_z$ in the $x$ and $z$ direction are varied. Hereafter, we use $N_x\times N_z$ denotes the grid resolutions, e.g. $32\times32$ representing $N_x=32$ and $N_z=32$. The grid numbers in the $y$ direction are set as $N_y=65$ for the coarser mesh cases ($21\times21$ and $32\times32$) and $N_y=97$ for the finer mesh cases ($63\times63$ and $84\times84$). Note that the CNN-K3 model examined here is the same as that tested in $\S$ \ref{sec:$posteriori$}, only DNS data at $Re_{\tau}=178$ is used for training, while this test is conducted at a higher Reynolds number.

\begin{figure}
  \centering
  {\includegraphics[width=0.475\textwidth]{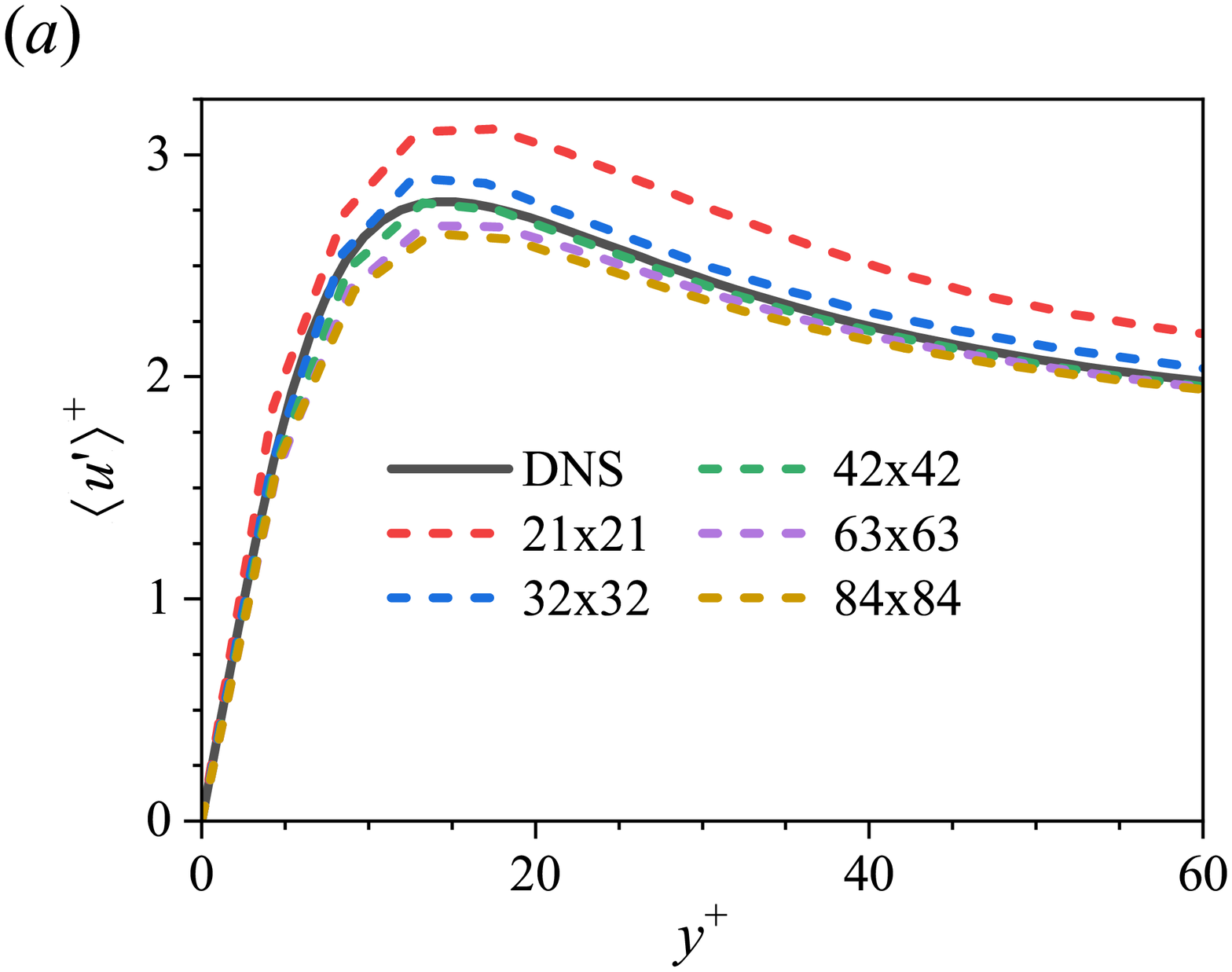}}
  {\includegraphics[width=0.475\textwidth]{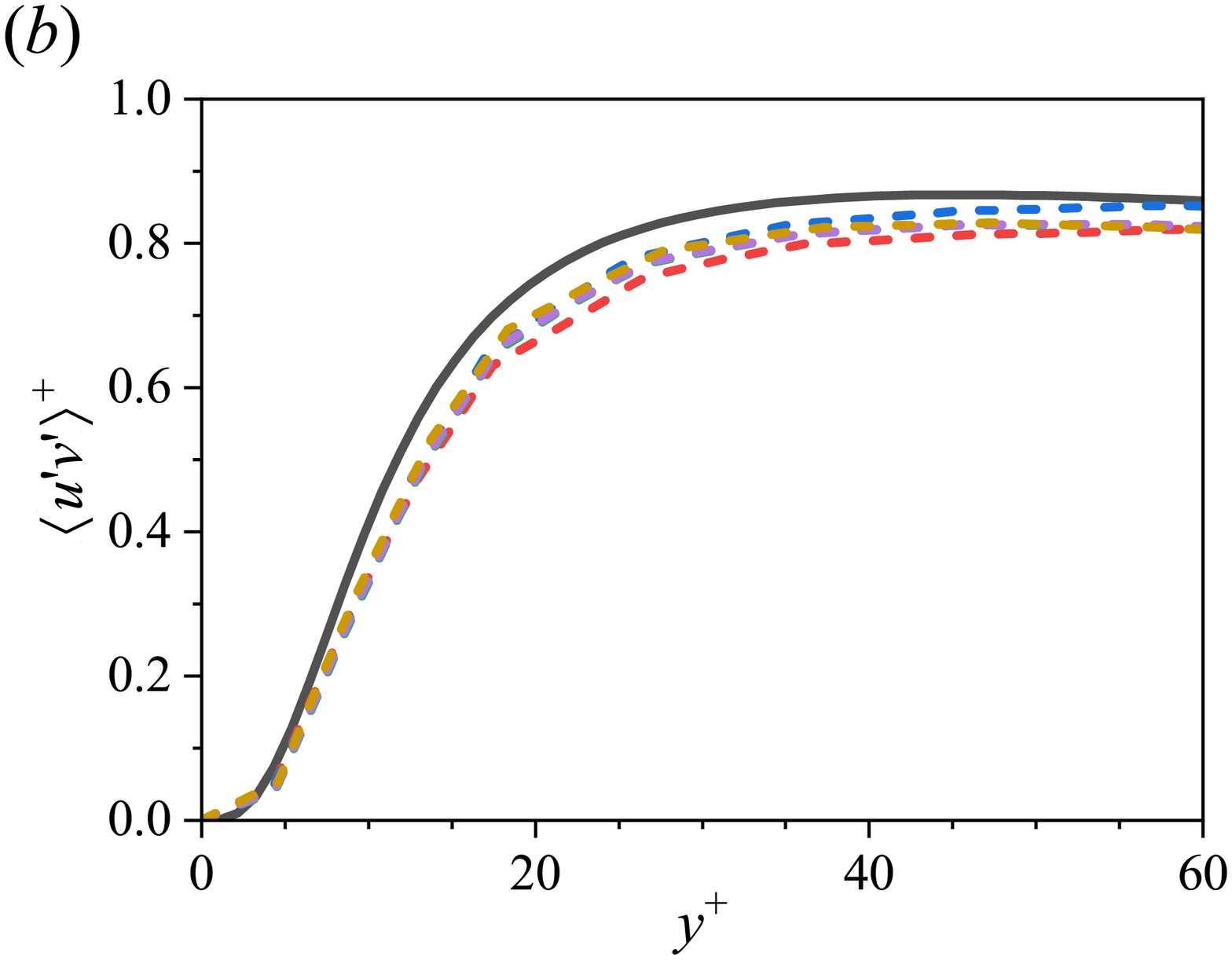}}
\caption{The r.m.s. velocity fluctuations (a) and Reynolds shear stress
(b) from LESs with CNN-K3 at $Re_{b}=16800$ in different
grid resolutions. All the LESs are conducted with domain size
$L_x$, $L_y$, $L_z=\pi$, 2, $\pi/2$.} \label{fig:resolution}
\end{figure}

As can be seen from figure \ref{fig:resolution}(a), the result from
the resolution $42\times 42$, which is the same as the resolution used in training, provides fairly well agreement with the DNS result. As the resolution becomes coarser (resolution
$32\times32$), the result of CNN-based model begins to deviate from the DNS result but is still in the rational range. When the resolution is as coarse as half of the resolution of training data
($21\times21$), the grid becomes too coarse for the CNN-based model
to generate a correct result, r.m.s. velocity fluctuations are dramatically overpredicted and the Reynolds shear stress is underestimated. Although not so accurate, the model still keeps numerical stability and does not diverge in such coarse resolution. On the other hand, the simulation in finer resolutions is more stable and authentic compared to the coarser ones, but a decline in performance is also clearly seen from the contrast of velocity fluctuations obtained in resolution $42\times42$, $63\times63$, and
$84\times84$. The difference is not obvious for the predicted Reynolds shear stresses as shown in figure \ref{fig:resolution}(b). The previous tests indicate the CNN-based models do not show mesh convergence trends as traditional models, i.e. a finer grid can not guarantee a better result. It results from that the ML-based models are trained with
data at specific filter size, the models prone to obtain better
results in the similar grid resolutions as training data rather than
finer ones or the coarser ones. Overall, the CNN-based model is able
to generate stable and reasonable solutions in a wide range of
resolutions, exhibiting excellent numerical robustness.

Finally, table \ref{tab:time} evaluates the models' complexity by comparing the number of parameters, multiply-accumulate (MAC) operations and the time consumption in realistic simulations of LES600 case (grid sizes $N_x, N_y, N_z=42, 97, 42$). The number of parameters determines the model size, while MAC denotes the number of calculations in a forward pass of the neural networks. The parameters numbers of CNN-K1 and ANN are quite small compared to the nonlocal models. CNN-K3 has about 30 times more parameters than CNN-K1. With the increase of kernel size, the number of parameters multiplies, so as the number of MAC operations. Note that the MACs for data-driven models (measured in billion) are much larger than the traditional models like SM, which means the models would be inefficient when working serially in practice and parallel running is necessary. When embedding these data-driven models into the realistic LESs, the discrepancy of time consumption for local models and nonlocal models is not so large as the MACs, since the convolution operation is well optimized by the mainstream machine learning libraries and parallel devices. Here, the ML algorithms and CFD simulations are performed on different devices (as introduced in $\S$ \ref{sec:hmlcfd}) of the same machine in the HML-CFD framework. The time consumption of LES without the SGS model (i.e. coarse DNS) under the same condition is utilized to norm other models’ time consumptions. Through the parallelization of ML algorithms, the two most accurate models in the $posteriori$ test, i.e. CNN-K3 and CNN-K5, merely spent $39\%$ and $44\%$ extra cost respectively, that are comparable to the traditional models. These results show the potential of ML-based models for replacing traditional models as an efficient tool in reality. The current simulations conducted here are small-scale, the HML-CFD framework will have a great advantage in large-scale computing.

\begin{table}
  \begin{center}
\def~{\hphantom{0}}
  \begin{tabular}{lcccccc}
  SGS model & No model & ANN & CNN-K1 & CNN-K3 & CNN-K5 & CNN-K7\\
  Parameters (K) & 0.00 & 1.08 & 0.98 & 30.37 & 84.13 & 164.77 \\
  MACs (G) & 0.00 & 0.17 & 0.17 & 5.2 & 14.40 & 28.19 \\
  Time consumption & 1.00 & 1.77 & 1.35 & 1.39 & 1.44 & 1.45 \\
  \end{tabular}
\caption{Comparison of model size and computing efficiency for the LESs with different models in LES600 case. The results of time consumption are scaled by the time required by LES with no model for comparison convenience. Here, the unit $K$ and $G$ denote $10^{3}$ and $10^{9}$ respectively.}
  \label{tab:time}
  \end{center}
\end{table}

\section{Conclusions}\label{sec:conclusion}

Based on the convolutional neural network, several SGS stress models are developed and examined in turbulent channel flow. CNN can automatically exploit flow features from raw flow variables, thus avoiding the trouble of feature selection. The naturally nonlocal property of CNN enables the prediction process accounting for spatial correlations in a wide flow region and contributes to the numerical stability in realistic LES. These nonlocal models were trained only with DNS data at $Re_{\tau}=178$, and tested in both $Re_{\tau}=178$ and $600$ case. In the $priori$ test, the correlation coefficients between $\tau_{ij}$ predicted by the nonlocal CNN-based models and $\tau^{fDNS}_{ij}$ is as large as around $0.87$, which is much larger than that of ANN. In the $posteriori$ test, the CNN-based model accurately predicted the mean flow and root-mean-square of velocity fluctuations, showing an advantage over ANN and traditional models in terms of accuracy. These nonlocal data-driven models also well predicted the backscatter without incurring numerical instability. The influence of filter kernel size (or receptive fields) of CNN is also investigated, the receptive field with appropriate size includes spatial information into the prediction process which is beneficial for the SGS modelling, while too large scope size is unnecessary. The two-point correlation analysis explained the impact of kernel size in a physical perspective, providing a quantitative method to guide the design of nonlocal models. On the other hand, the CNN-based model is well fitted to different grid resolutions and shows excellent numerical robustness. Due to our HML-CFD framework, the CNN-based model can be as efficient as traditional models in practical simulations. This framework is suitable for various machine learning algorithms and physical problems, not limited to the LES modelling problem.

Although CNN has the advantage in feature extraction and information perception, it still has some shortcomings. One is that the CNN can be readily applied on a structured grid but suffers in configuration with complex geometry or unstructured grids because traditional convolutional kernels can not contain unstructured topology information. Data transformations or specially designed convolutional filters may be helpful for the solution of this problem. The ANN-based model is more flexible in this aspect since it works in a point-to-point manner. Besides, we have successfully carried out the extrapolation test in the Reynolds numbers and grid resolutions different from those of training data, but it is still a challenge for ML-based models to extrapolate to different flow types that are not contained in the training process. Interpretability is also a big issue for all data-driven approaches, which will be a target in our future work.

\section*{Acknowledgements}

The authors are very grateful to Profs. N.-S. Liu and Z.-H. Wan at USTC for many useful discussions on the numerical results. This work was supported by Natural Science Foundation of China (Nos. 92052301 and 11621202) and Science Challenge Project (No. TZ2016001). The present direct numerical simulations were performed on the supercomputing system in the Supercomputing Center of USTC.

\bibliographystyle{jfm}
\bibliography{jfm}

\end{document}